\documentclass{article}[12pt]
\usepackage{graphics}
\usepackage{epsfig}
\usepackage{epstopdf}
\usepackage{amsmath}
\usepackage{array}
\begin{document}
\begin{center}
{\LARGE Holographic equipartition and the maximization of entropy \\[0.2in]}
{ Krishna P B and Titus K Mathew\\
e-mail:krishnapb@cusat.ac.in, titus@cusat.ac.in \\ Department of
Physics, Cochin University of Science and Technology, Kochi, India.}
\end{center}

\begin{abstract}
The accelerated expansion of the universe can be interpreted as a tendency to satisfy the holographic equipartition. It can be
expressed by  a simple law,
$\Delta V = \Delta t\left(N_{surf}-\epsilon N_{bulk}\right),$ where $V$ is the Hubble volume in Plank units, $t$ is the cosmic
time plank units and $N_{surf/bulk}$ is the degrees of freedom on the horizon/bulk of the universe. We show that this holographic
equipartition law effectively implies the maximization of entropy. In the cosmological context, a system that obeys the
holographic equipartition law behaves as an ordinary macroscopic system that proceeds to an equilibrium state of maximum entropy.
We consider the standard $\Lambda$CDM model of the universe and have shown that it is consistent with the holographic
equipartition law. Analyzing the entropy evolution we find that it also proceeds to an equilibrium state of
 maximum entropy.
\end{abstract}



\section{Introduction}
\label{intro}The connection  between gravity and thermodynamics is a recently emerged area of research. This connection was first
demonstrated by Bekenstein and Hawking in the context of black hole mechanics\cite{Bekenstein1,Bekenstein2,Hawking1,Hawking2}.
In1995 Jacobson\cite{Jacob1}, using the Bakenstein law\cite{Bekenstein1} of horizon entropy,
showed that the Einstein field equations of gravity can be obtained from the thermodynamical
principles. Possible schemes for relating  gravity and thermodynamics have been discussed for a variety of gravity theories: see \cite{Eling1}
and \cite{Paddy1} and references therein. In 2011 Verlinde\cite{verlinde} proposed a model where gravity is treated as an emerging
phenomenon, contrary to our common perception. He considered gravity as an entropic force caused by the entropic gradients
due to the changes in the distribution of the material bodies. Interpreting gravity as a tendency to maximize entropy, he derived
Newton's law of gravitation using holographic principle and equipartition law of energy. A similar proposal was made by
Padmanabhan \cite{Paddy2}, and he derived the Newton's law of gravity  by combining the equipartition law of energy and
the thermodynamic relation $S=E/2T$. Here $S$ and $T$ are the entropy and temperature of the horizon  and
E represents the active gravitational mass.

In a recent work Padmanabhan\cite{Paddy3} analyzed the possibility of describing spacetime itself as an emergent structure.
In this approach it has to be noted, first of all, there is a conceptual difficulty in treating time as having emerged from
some pre-geometric variables. Also any emergent description of the gravitational fields around finite gravitating systems
has to consider the space around them as pre-existing. Hence, for the finite gravitating systems without a special status for the
time variable, a consistent formulation of  spacetime itself as an emergent structure is too difficult . But in the study of
cosmology, we can overcome these difficulties by choosing the time variable as the proper time of the geodesic
observers to whom CMBR appears homogeneous and isotropic. This provides a strong justification for describing the evolution
of the universe as the emergence of cosmic space with the progress of cosmic time.

To justify this approach let us consider  a pure de Sitter universe with a constant Hubble parameter which
obeys the holographic principle in the form $N_{surf} = N_{bulk},$ where $N_{surf}$ is the number of degrees of freedom on
the boundary surface and $N_{bulk}$ is the number of degrees of freedom in the bulk. This condition relates the degrees of
freedom in the bulk, determined by the equipartition condition, to the degrees of freedom on the boundary surface and thus
can be called as the holographic equipartition. Even though  our universe is not exactly de Sitter there is considerable evidence
indicating that it is proceeding towards a pure de Sitter state. Based on the above facts, it has been suggested that the expansion
of the universe or  equivalently the emergence of the cosmic space is driven by the departure from the holographic equipartition.
Following Padmanabhan's work Cai \cite{Cai1}deduced the Friedmann equation for a higher dimensional FLRW universe
using the holographic equipartition law. He also obtained the corresponding dynamical equations of the
universe in Gauss-Bonnet gravity and in more general Lovelock gravity models. Role of holographic equipartition in
the case of the collapse of gravitating systems were analyzed in reference\cite{Cao1}.

 It is well known that, ordinary, macroscopic systems evolve to a state of maximum entropy consistent with their constraints\cite
{Callen1}.
Such a maximum entropy state is not achievable in gravity-dominated systems unless the divergence of entropy
is prevented by the formation of a black hole\cite{Bekenstein2,Bekenstein3,Sewell1}. However, in recent work Pavon and Radicella
 \cite{Diego1}have shown that a Friedmann universe with a Hubble expansion history can behave as an ordinary macroscopic system 
with the entropy tending to some maximum value. As we have discussed above, it is possible to describe  the evolution of the
universe as being driven by the departure from the holographic equipartition and the universe is proceeding to a state that
satisfies the holographic equipartition. In this context it is of interest to analyze  the possibility that a system which is
consistent with holographic equipartition  behaves as an ordinary macroscopic system.

In this paper, we study the problem whether the holographic equipartition explicitly
implies the maximization of entropy or not. Our analysis shows holographic equipartition and entropy maximization are
equivalent. We  also check the consistency of the standard $\Lambda$CDM model both with the holographic equipartition and
the maximum entropy principle. The paper is organized as follows. In section 2, we discuss
the connection between the holographic equipartition law and maximum entropy principle. In section 3, we discuss the
consistency of the $\Lambda$CDM model both with the holographic equipartition law and maximum entropy
principle. In section 4, we present our conclusions.

\section{Holographic equipartition and maximization of entropy}
\label{sec:entro}

In this section we present a detailed analysis of the holographic equipartition principle and its implications.
This is followed by a detailed discussion of  the maximum entropy principle according to which the universe behaves
as an ordinary macroscopic system  that approaches an equilibrium state of maximum entropy. We, then, show that these two
 principles are  equivalent to one another.

\subsection{Holographic equipartition}
Let us start with a brief description of Padmanabhan's idea of holographic equipartition.
The holographic equipartition condition can be stated as,
\begin{equation}
 N_{surf} = N_{bulk}
\end{equation}
where $N_{surf}$ is the degrees of freedom on the surface of the horizon and $N_{bulk}$ is the degrees of freedom in
the bulk cosmic fluid enclosed by the horizon. This condition is exactly satisfied by a pure de Sitter universe.
In calculating the $N_{surf},$ one can take either Hubble horizon or event horizon as the boundary of the universe.
For simplicity the Hubble horizon having radius $r \sim \frac{1}{H},$ can be taken as the boundary surface.
Plank area, $L^2_P\sim G$ is equivalent to one degree of
freedom on the boundary surface, where $L_P$ is the Plank length. The degrees of freedom on the Hubble horizon can
then be defined as,
\begin{equation}\label{eqn:Nsurf1}
 N_{surf} = {4 \pi \over L^2_P H^2}.
\end{equation}
The effective number of the bulk degrees of freedom inside the Hubble horizon at temperature $T$ is given by
\begin{equation}
N_{bulk} = {|E|\over{1\over2} {k_B T }}.
\end{equation}
We shall now assume the existence of thermal equilibrium between the bulk and the horizon. Hence, the most natural choice
of the horizon temperature, the Gibbons-Hawking temperature, $T = {H\over 2\pi},$ can also be taken as the temperature
of the bulk which has already being emerged as per the emergence paradigm.
$E$ in the above equation represents the total energy enclosed by the horizon and can be taken as the
Komar energy $|\rho+3p|V,$
where $V={4\pi\over 3 H^3},$ is the volume
enclosed by the Hubble horizon. So the expression for the bulk degrees of freedom becomes
\begin{equation} \label{eqn:Nbulk1}
 N_{bulk} = -\epsilon{2(\rho+3p) V \over k_B T }
\end{equation}
where $\epsilon = +1 $, if $ (\rho+3p)< 0 $ as for dark energy and $\epsilon = -1$,  if $(\rho+3p)>0$
as for normal matter and radiation and consequently the degrees of freedom is
always a positive number. The equation $N_{surf}=N_{bulk}$ can also
be written as $|E|={{1\over2}{k_B T N_{surf}}}$. This condition
relates the the degrees of freedom in the bulk which is
determined by the equipartition condition to the degrees of freedom on the horizon. Hence it can be referred to as the
`holographic equipartition'. Even though our universe is not exactly
 de Sitter at present, it is proceeding to a pure de Sitter phase that satisfies the holographic equipartition.
 Hence it may be argued  that the accelerated expansion of the universe, or equivalently the emergence of space,
 is due to the quest of the universe for satisfying the holographic equipartition. This means that the expansion of the
 universe is driven by the difference between the number of degrees of freedom on the horizon and that in the bulk.
 Mathematically this can be expressed as
\begin{equation}
 \Delta V={\Delta t(N_{surf}-\epsilon N_{bulk})}
\end{equation}
where $V$ is the Hubble volume in Planck units and $t$ is the cosmic time in Planck units. In an infinitesimal cosmic time
interval $dt$, the change in Hubble volume $dV$ can be expressed as
\begin{equation} \label{eqn:dVdt1}
 {dV\over dt} ={L^2_P(N_{surf}- \epsilon N_{bulk})}.
\end{equation}
It is possible to obtain the Friedmann equation from  this postulate\cite{Paddy3}. For a universe consisting of non-relativistic
 matter, radiation and dark energy, the above equation becomes
\begin{equation} \label{eqn:dVdt3}
 \frac{dV}{dt}=L^2_P \left(N_{surf}+N_{matter}+N_{rad} - N_{de} \right),
\end{equation}
where $N_{matter}$ is the degrees of freedom corresponding to non-relativistic matter, $N_{rad}$ is the degrees of freedom
corresponding to radiation, $N_{de}$ is that of dark energy and all of them are positive definite.
As the universe attains holographic equipartition, the derivative
${dV/dt} \to 0,$ which inevitably demands the presence of dark energy in the universe.

We will now express the left hand side of equation (\ref{eqn:dVdt3}) in terms of the deceleration parameter and
discuss its implications. The time evolution of the Hubble volume depends on the time evolution of the Hubble parameter.
Using the expression of the Hubble volume, the time derivative of the volume can be written as,
\begin{equation} \label{eqn:dVdt12}
 \frac{dV}{dt}=-\frac{4\pi}{H^2} \left(\frac{\dot H}{H^2} \right).
\end{equation}
From the basic definition of the deceleration parameter $q=-1-(\dot H / H^2)$ , the above equation becomes,
\begin{equation} \label{eqn:dVdt2}
 \frac{dV}{dt}=\frac{4\pi}{H^2} \left(1+q\right).
\end{equation}
This equation shows that for $dV/dt \geq 0,$ the deceleration parameter satisfies, $\left(1+q\right)\geq0$. The
equality sign holds for $q=-1$ which corresponds to de Sitter universe. Since our universe approaches a de Sitter
phase in the long run, the rate of change of the Hubble volume, $dV/dt \to 0$ asymptotically.

Combining the equations (\ref{eqn:dVdt1}) and (\ref{eqn:dVdt2}) we get,
\begin{equation}
 \frac{4\pi}{H^2} \left(1+q\right) = L^2_P \left(N_{surf} - \epsilon N_{bulk} \right).
\end{equation}
from which we obtain the expression for the deceleration parameter $q$ in terms of the degrees of freedom as,
\begin{equation} \label{eqn:q3}
 q = -\epsilon \left(\frac{N_{bulk}}{N_{surf}} \right).
\end{equation}
The evolution of deceleration parameter is thus equivalent to the evolution of the ratio of the
degrees of freedom in the bulk and in the horizon. Substituting the expression for $N_{bulk}$ from equation
(\ref{eqn:Nbulk1}), the above expression becomes,
\begin{equation}
 q = +\epsilon^2 \left({2 (\rho+3 p) V \over k_B T} \right)  \frac{1}{N_{surf}},
\end{equation}
where $\epsilon^2=+1.$ For  non-relativistic matter or radiation $(\rho+3 p) >0,$ and so
$q >0;$ but for dark energy $(\rho+3 p) <0$ and hence $q<0.$ Substituting for $N_{surf}$ from equation (\ref{eqn:Nsurf1})
and by taking $k_B=1$ the above equation becomes,
\begin{equation} \label{eqn:q2}
 q = \frac{4 \pi L^2_P}{3H^2} \left(\rho + 3 p \right).
\end{equation}

Assuming the barotropic equation, $p=\omega \rho,$ where $\omega$ is the equation of state constant of the cosmic
fluid, the $q$ parameter equation takes the form,
\begin{equation}
 q = \frac{4 \pi L^2_P}{3H^2} \left(1 + 3 \omega \right) \rho.
\end{equation}
For the radiation dominated phase of the universe, having the equation of state with $\omega \sim 1/3,$  the Friedmann
equation, $3H^2 \sim  8\pi G \rho_{\gamma},$ gives the deceleration parameter  $q \sim 1.$
Then from equation (\ref{eqn:q3})we have
\begin{equation}
 N_{surf}=-\epsilon N_{bulk}.
\end{equation}
Since $\epsilon =-1$ for radiation component, $N_{surf}=N_{bulk}.$
 The degrees of freedom in both the bulk and horizon surface are equal in this case. But the rate of change of Hubble volume,
$dV/dt \propto 2 N_{surf}.$  So, even though the degrees of freedom are equal in the radiation
dominated phase, it does not lead to the condition $dV/dt=0,$ which is the equilibrium condition corresponding  to maximum entropy.

For the matter dominated phase of the universe, with equation of state having $\omega \sim 0,$  the  deceleration parameter
$q \sim 1/2.$ and hence according to equation (\ref{eqn:q3}),
\begin{equation}
 N_{surf}=-2\epsilon N_{bulk}.
\end{equation}
Since $\epsilon=-1$ for matter, $N_{surf}=2 N_{bulk},$ holographic equipartition is not satisfied.
The rate of change of Hubble volume is then, $dV/dt \propto (3/2) N_{surf}.$

During the later stage of the evolution, when dark energy dominate over
other components,  the equation of state $\omega \sim -1$ where the cosmological constant is taken as
the dark energy component. The deceleration parameter, now, becomes, $q \sim -1.$ Therefore according to equation (\ref{eqn:q3}),
\begin{equation}\label{eqn:Nsb}
 N_{surf}=\epsilon N_{bulk}.
\end{equation}
Since in this case $\epsilon =+1,$ $N_{surf}=N_{bulk}.$
But unlike in the case of the radiation dominated phase,  the change in Hubble volume, $dV/dt = 0.$ So only in
the presence of dark energy, the equality of the degrees of freedom leads to the condition $dV/dt=0.$
This means  the condition $N_{surf}=N_{bulk}$ leads to $dV/dt \to 0$ only in the accelerated phase of the universe.
So, in general, we can write $N_{surf}=\epsilon N_{bulk}$ as statement of the holographic equipartition.

\subsection{Evolution of the universe and maximization of entropy}

It is well known that any isolated macroscopic system evolves to a state of maximum entropy called the
equilibrium state\cite{Callen1}. This  means that  the first derivative of the entropy, $S$ is always greater than zero and the
second derivative of it is less than zero in the long run
\begin{equation}\label{eqn:conditions1}
 \dot S \geq 0, \, always;  \, \, \, \, \, \,  \ddot S <0 \, \, at \, least \, in \, long \, run.
\end{equation}
 The derivative of the entropy can be taken
with respect to cosmic time or any other relevant variable. In the context of Newtonian gravity it has been shown that the
entropy of a gravitating system may diverge asymptotically, without attaining a maximum entropy state\cite{Lynden1}. But
when Newtonian theory is replaced by Einstein's gravity theory, such a catastrophe need not occur. In general relativity
the divergence of entropy is prevented by the formation of a black hole. Even if the black hole evaporates,
the total entropy, comprising the entropy of the black hole and emitted radiation, is on  the increase. It is possible
to achieve an equilibrium condition between the evaporating black hole and the radiation and as a result the system will
tend to a maximum entropy state \cite{Thorn1}. The evolution of the FLRW universe also shows the same thermal behavior;
it evolves towards a state of maximum entropy.

If our universe is evolving as an ordinary macroscopic system, it must proceed to a state of maximum entropy. More precisely
its entropy should not decrease, i.e, $\dot S \geq 0$ always and it must attain a state of maximum entropy in the long run
 i.e. $\ddot S < 0$ at least in the last stages of the evolution of the universe. This has been discussed in reference
\cite{Diego1} and the authors have shown that our universe is evolving towards a maximum entropy state. Let us, now, consider
the total entropy of the universe, which is the sum of the entropies of various constituents like the horizon, super massive
black holes, non-relativistic matter, radiation, etc. Compared to the horizon entropy, entropy of other components of
the universe are negligibly small, for instance the entropy contribution from super massive black holes is around 18 orders
smaller, and radiation entropy is 33 orders smaller and so on\cite{Egan1}. Thus the total entropy is approximately
equal to the horizon entropy. According to Bekenstein's result, the horizon entropy is \cite{Bekenstein1,Hawking1},
\begin {equation}
 S_H = \frac{A_H}{4},
\end{equation}
where $A_H$ is the area of the horizon. In the present case where we consider the Hubble horizon, with horizon radius
 $r_H \sim 1/H$  the rate of change of entropy is given by \begin{equation}\label{eqn:sdot1}
 \dot{S}_H = -\frac{2 \pi \dot H}{H^3}.
\end{equation}
Since $H>0$ always for an expanding universe, the horizon entropy will always increase, if $\dot H <0.$
In terms of the deceleration parameter  $-\dot{ H}/H^2 = 1+q$ and  as long as  $(1+q) \geq 0$, the rate of change of
Hubble parameter $\dot H <0.$ This relation guarantees that, the horizon entropy will never decrease. The observational
data on Hubble parameter also indicates that $\dot H<0.$ Numerical simulations\cite{Craw1,Carva1}, using the
observational data on Hubble parameter\cite{Simon1,Stern1}, have confirmed the above conclusion.

The attainment of equilibrium requires that $\ddot{S}_H <0$ at least in the long run. The $\ddot{S}_H$ can
be obtained using the expressions given above,
\begin{equation}\label{eqn:ddotS12}
 \ddot{S}_H= 2\pi \left[3 \left(\frac{\dot H}{H^2}\right)^2 - \left(\frac{\ddot H}{H^3} \right) \right].
\end{equation}
Hence the maximization of entropy demands,
\begin{equation} \label{eqn:ineq2}
 3 \left(\frac{\dot H}{H^2}\right)^2 < \left(\frac{\ddot H}{H^3} \right),
\end{equation}
That is,
\begin{equation}\label{eqn:ineq1}
 3 \left(1+q\right)^2 < \left(\frac{\ddot H}{H^3} \right).
\end{equation}
In the asymptotic limit, $q \to -1,$  the left hand side of the above inequality  tends to zero. But $\ddot H > 0$
always, as per the analysis of the observational data on Hubble parameter\cite{Craw1,Carva1,Simon1,Stern1}. Hence the
inequality in equation (\ref{eqn:ineq1}) is satisfied by the expanding universe at later stages and consequently
the entropy of the universe  gets  saturated in the asymptotic limit.

\subsection{Holographic equipartition vs entropy maximization}

We have already noted the accelerated expansion of the universe is due to the difference
in the number of degrees of freedom on the surface and that in the bulk.
Our universe, also, behaves as an ordinary macroscopic system that obeys the conditions in equation (\ref{eqn:conditions1}).
In the previous section we have seen that these two will be satisfied, if $\dot H <0$ and inequality
(\ref{eqn:ineq2}) is satisfied.

In this section we  show that the holographic equipartition effectively implies the maximization of entropy.
Combining equations (\ref{eqn:dVdt1}) and
(\ref{eqn:dVdt12})
we have
\begin{equation}
 -\frac{4\pi}{H^2} \left(\frac{\dot H}{H^2} \right) = L_p^2 \left(N_{surf} - \epsilon N_{bulk} \right).
\end{equation}
Using equation (\ref{eqn:sdot1}), we can write the rate of change of entropy as,
\begin{equation} \label{eqn:sdot2}
 \dot S = \frac{2\pi}{H} \left(1-\epsilon \frac{N_{bulk}}{N_{surf}} \right).
\end{equation}
For $\dot S \geq 0,$ the condition is
\begin{equation}
 1-\epsilon \frac{N_{bulk}}{N_{surf}} \geq 0
\end{equation}
or equivalently  $(N_{surf}-\epsilon N_{bulk}) \geq 0$ always or as per equation (\ref{eqn:dVdt1}),
$dV/dt \geq 0$ .This, in turn, implies that
$\epsilon N_{bulk}$ never exceeds $N_{surf}.$  In the asymptotic limit when the system attains equilibrium,
$\dot S \to 0,$ we will have $N_{surf}=\epsilon N_{bulk},$ which is exactly equation (\ref{eqn:Nsb}) obtained previously
for the dark energy dominating case.
From equations (\ref{eqn:q3}) and (\ref{eqn:sdot2}), we can write
\begin{equation}
 \dot S = \frac{2\pi}{H} \left(1+q \right),
\end{equation}
where $S$ is the total entropy of universe, approximately same as $S_H.$ Since the minimum value of $q = -1$ is
attained only in the asymptotic stage the total entropy will never decrease.

Taking the derivative of equation (\ref{eqn:sdot2}) with respect to cosmic time, we get,
\begin{equation}
 \ddot S = -\frac{2\pi\dot H}{H^2} \left( 1- \epsilon \frac{N_{bulk}}{N_{surf}} \right) +
              \frac{2\pi}{H} \frac{d}{dt} \left(1-\epsilon \frac{N_{bulk}}{N_{surf}} \right).
\end{equation}
Substituting  $-\dot H /H^2 = 1-\epsilon \frac{N_{bulk}}{N_{surf}},$ the above equation becomes,
\begin{equation}
 \ddot S = 2\pi \left( 1- \epsilon \frac{N_{bulk}}{N_{surf}} \right)^2 +
              \frac{2\pi}{H} \frac{d}{dt} \left(1-\epsilon \frac{N_{bulk}}{N_{surf}} \right).
\end{equation}
The first term in the above equation is always positive and it becomes zero asymptotically, while the second term becomes
negative in the long run. Hence for $\ddot S < 0,$ the condition to be satisfied is,
\begin{equation}
\left|\frac{1}{H} \frac{d}{dt}\left(1-\epsilon \frac{N_{bulk}}{N_{surf}} \right) \right| > \left(1-\epsilon
\frac{N_{bulk}}{N_{surf}} \right)^2.
\end{equation}
Using equation (\ref{eqn:q3}) the above condition can be written as,
\begin{equation}
\left|\frac{1}{H} \frac{d}{dt}\left(1 + q \right) \right| > \left(1+q \right)^2.
\end{equation}
In the dark energy dominated phase where $N_{surf}=\epsilon N_{bulk},$ the right hand side of the above inequality becomes zero.
In the radiation dominated phase $N_{surf} = -\epsilon N_{bulk},$ the right side will be greater than zero. Hence the
inequality in general will be satisfied only in a dark energy dominated phase. Hence, only in the dark energy dominated phase we
can guarantee the non-positivity of $\ddot S.$ In the radiation dominated phase even though $N_{surf}=N_{bulk}$ the entropy
can grow without any bound.

If expression for the deceleration parameter, $q = -1 -\dot H/H^2,$ is substituted in the above inequality condition, it becomes,
\begin{equation}
 3 \left(\frac{\dot H}{H^2}\right)^2 < \left(\frac{\ddot H}{H^3} \right),
\end{equation}
This is exactly the same as equation (\ref{eqn:ineq2})which is the condition for entropy maximization. This proves the
equivalence of  the holographic equipartition and maximization of entropy .

\section{An analysis based on the standard $\Lambda$CDM model}

In this section we analyze  the consistency of the standard $\Lambda$CDM model with the holographic equipartition
and the maximum entropy principle.  Our aim here is to check whether this standard model of the
universe supports our earlier arguments regarding the equivalency of the holographic equipartition law and the maximization
of entropy.
As is well known, the $\Lambda$CDM model with cold dark matter and cosmological constant as the major components
of the universe is the standard model of the universe. This model is in good agreement with observational data except with some
discrepancies regarding the nature and evolution of the cosmological constant. It fails to explain the current value of the
cosmological constant and also the equality in the orders of the densities of non-relativistic matter  and dark energy.
However, in almost all other respects, this model is considered to be the most successful one.

Considering a flat FLRW universe with non-relativistic matter and cosmological constant as the cosmic components, the Friedmann
equation can be written as,
\begin{equation}
  H^2 = \frac{8\pi G\rho_m}{3} + \frac{\Lambda}{3},
\end{equation}
where $\rho_m$ is the density of the non-relativistic natter and $\Lambda$ is the cosmological constant. The scale factor of the expanding universe
can be obtained as\cite{TKM1},
\begin{equation}\label{eqn:a}
 a(t)= \left(\frac{\Omega_{mo}}{\Omega_{\Lambda}}\right)^{1/3} \sinh^{2/3} \left(\frac{3}{2} \sqrt{\Omega_{\Lambda}} H_0 t \right),
\end{equation}
where $\Omega_{mo}=8\pi G \rho_{mo}/3H_0^2$ is the present value of the mass density parameter with $\rho_{m0}$ as the present
matter density, $\Omega_{\Lambda}=\Lambda /3H_0^2$ and $H_0$ is the present value of the Hubble parameter. This expression for the
scale factor has desirable  behavior both at initial stage and at the later stage. For instance, as $t \to 0$ the scale
factor, $a(t) \to t^{2/3}$ corresponds to the matter dominated era, while at $t \to \infty$ the scale factor,
$a(t) \to \exp(\sqrt{\frac{\Lambda}{3}} \, t),$ corresponds to the de Sitter phase. This means, during the evolution,
the universe has undergone a transition from a matter dominated phase where the expansion was decelerated to a dark
energy dominated phase where expansion is accelerating. The corresponding Hubble parameter is,
\begin{equation}\label{eqn:H}
 H=H_0 \sqrt{\Omega_{\Lambda}} \coth \left(\frac{3}{2} \sqrt{\Omega_{\Lambda}} H_0 t \right).
\end{equation}
The evolution of the Hubble parameter is consistent with the evolution of the scale factor, such that as $t \to 0,$ the Hubble
parameter $H \to 2/3t$ and at $t \to \infty$ the Hubble parameter $H \to \sqrt{\Lambda /3}.$

We shall first check the consistency of this model with holographic equipartition law.
 As explained in section \ref{sec:entro}, FLRW universe satisfies the relation given by equation (\ref{eqn:dVdt1}), if it
evolves to a state that satisfies holographic equipartition.
The number of degrees of freedom on the boundary surface is given by equation (\ref{eqn:Nsurf1}). The number of degrees
of freedom in the bulk region of space is the sum of the degrees of freedom associated with matter and the degrees of
freedom  associated with the dark energy.

 Let us now calculate the degrees of freedom associated with the matter distribution. For this pressureless, non-relativistic matter,
 $(\rho_m+3p_m)>0$, where $\rho_m$ is the  matter density and pressure, $p_m = 0.$ Hence we take $\epsilon=-1$ to make
 $N_{matter} $ positive. Then from equation (\ref{eqn:Nbulk1}), the  degrees of freedom corresponding to the non-relativistic
 matter is obtained as,
 \begin{equation} \label{eqn:Nmatter}
 N_{matter} = {2 \rho_m V \over k_B T},
\end{equation}
Using the energy conservation equation for matter $d(\rho_m a^3) =-p_m da^3$ \cite{Paddy3}, or the continuity equation
$\dot\rho +3H(\rho+p)$ \cite{Cai1}, one gets the matter density as $\rho_m =\frac{\rho_{m0} a^{-3}}{a_0^{-3}}$, where
$\rho_{mo}= \frac{3H_0^2\Omega_{mo}}{8\pi G}$ is the present matter density
and we set the present value of the scale factor as, $a_0=1.$

We will now calculate the degrees of freedom  associated with the dark energy. In the standard $\Lambda$CDM model,
the dark energy density $\rho_{\Lambda}=\Lambda/8\pi G$ and the pressure $p_{\Lambda} =-\rho_{\Lambda}$. Hence for dark
energy,  $(\rho_{\Lambda}+3p_{\Lambda})<0$ and we take
$\epsilon=1$ to make $N_{de}$ positive. Then, the degrees of freedom associate with the dark energy can be obtained from
equation (\ref{eqn:Nbulk1}) as,
\begin{equation} \label{eqn:Nde}
 N_{de} = {4\rho_{\Lambda} V \over k_B T}.
\end{equation}
Hence the total bulk degrees of freedom is,
\begin{equation}\label{eqn:Nbulk12}
 N_{bulk}={2 \rho_m V \over k_B T} + {4\rho_{\Lambda} V \over k_B T}
\end{equation}

Now, equation (\ref{eqn:dVdt1}) can be written as,
\begin{equation}
 \frac{dV}{dt}= L_p^2 \left(\frac{4\pi}{L_p^2 H^2} + \frac{2\rho_{m0} a^{-3} V}{k_B T} - \frac{4\rho_{\Lambda} V}{k_B T}
 \right)
\end{equation}
 Using equation (\ref{eqn:dVdt12}) and by taking $T=H/2\pi,$ the Gibbons-Hawking
temperature, the above equation becomes
\begin{equation}\label{eqn:RL}
 -{4\pi \dot H \over H^4 } = \frac{4\pi}{H^2} + \frac{2\pi L_P^2 H_0^2}{k_B G H^4} \left(\Omega_{m0} a^{-3} -
 2\Omega_{\Lambda} \right).
\end{equation}
Using equation (\ref{eqn:H}), the left hand side of the above equation can be written as,
\begin{equation}
 \frac{-4\pi \dot H}{H^4} = \frac{6\pi}{H_0^2 \Omega_{\Lambda}} \tanh^2 \left(\frac{3}{2} \sqrt{\Omega_{\Lambda}} H_0 t \right)
 sech^2 \left(\frac{3}{2} \sqrt{\Omega_{\Lambda}} H_0 t \right).
\end{equation}
Taking $L_P^2 \sim G, \, k_B=1$ and by using equations (\ref{eqn:a}) and (\ref{eqn:H}) the right hand side of
equation (\ref{eqn:RL}) reduces to the left hand side. This shows that the expansion of the Hubble sphere in the
$\Lambda$CDM model is driven by the departure from the holographic equipartition.

\begin{figure}[h]
\centering
 \includegraphics[width=0.45\textwidth]{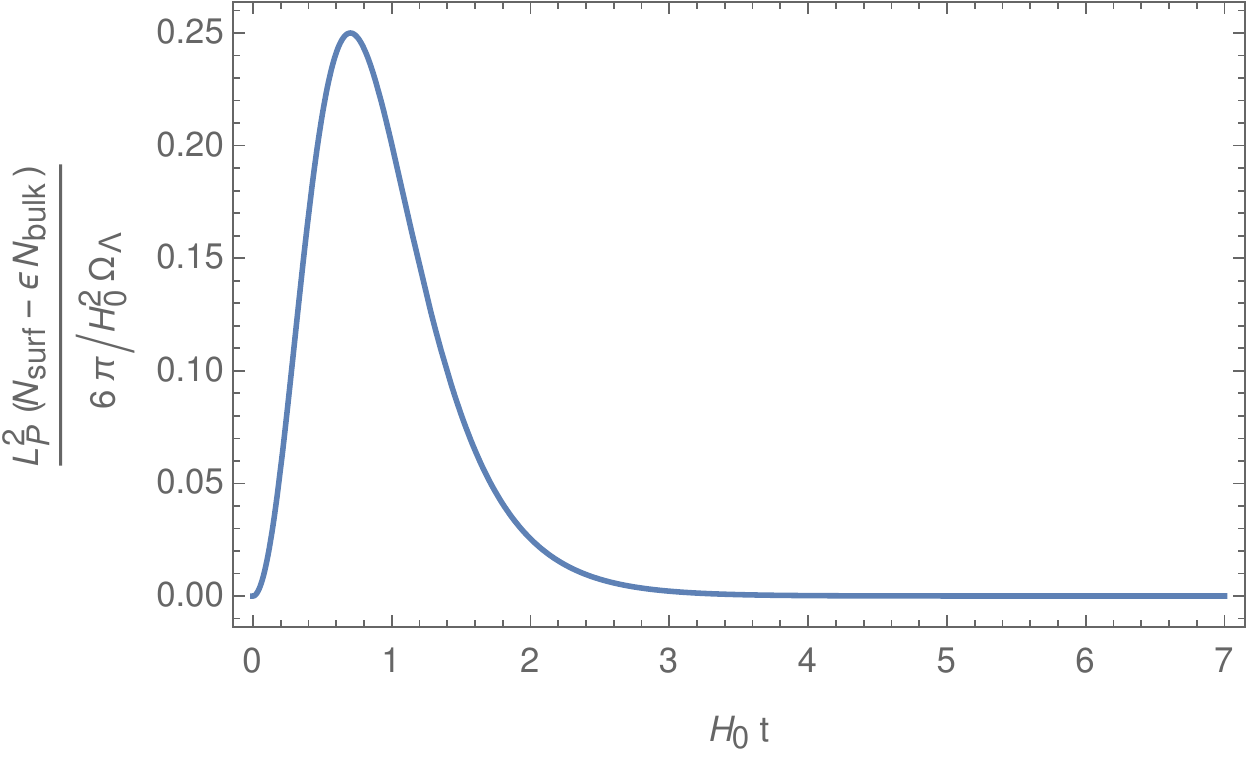}
 \caption{Evolution of the holographic discrepancy with the progress in cosmic time}
 \label{fig:RL}
\end{figure}

Now let us check whether this universe evolves to a state that satisfies the holographic equipartition condition
$N_{surf}=N_{bulk}.$ The number of degrees of freedom on the surface,
\begin{equation}
 N_{surf} = \frac{12\pi }{L_P^2 \Lambda}
\end{equation}
as $t \to \infty$, where we have used the expression for the Hubble parameter from equation (\ref{eqn:H}). Since the dark
energy is dominating over the matter density in the asymptotic limit,
only second term  in equation (\ref{eqn:Nbulk12})
survives as $\rho_m \to 0$ when $t \to \infty.$ Then 
the bulk degrees of freedom becomes,
\begin{equation}
 N_{bulk} = \frac{4\rho_{\Lambda} V}{k_B T}.
\end{equation}
Substituting for $\rho_{\Lambda}, V$ and $T$ as earlier, it can be shown that $N_{bulk}$
ultimately become equal to $N_{surf}$ in the asymptotic limit. We have plotted the holographic discrepancy
$N_{surf} - \epsilon N_{bulk} $ against the cosmic time $H_0 t$ in figure \ref{fig:RL}. The initial increase in the
holographic discrepancy is due to the deceleration in the early period and it attains the maximum value at the end
of this phase. This discrepancy starts to decrease when the universe makes its transit into the accelerating phase
and approaches zero in the asymptotic limit when $N_{surf}=\epsilon N_{bulk}.$ This shows the consistency of $\Lambda$CDM
model with the holographic equipartition law.

According to our earlier findings, a system that satisfies holographic equipartition behaves as an ordinary macroscopic system
that proceeds to an equilibrium state of maximum entropy. Now we will provide further support for our argument by analyzing the entropy
evolution of the $\Lambda$CDM model.

 The total entropy of the matter dominated Friedmann universe can be approximated as the sum of the entropy of matter inside
 the cosmic horizon and the entropy of the horizon itself i.e,
\begin{equation}
 S=S_H+S_m
\end{equation}
where $S_H$ and $S_m$ are the entropy of horizon and entropy of matter respectively.  But it is known that the matter
entropy is much less than the horizon entropy. It lags 35 orders of magnitude behind the horizon entropy \cite{Egan1}.
Hence the total entropy of the universe can be approximated as the horizon entropy \cite{Diego1}.
i.e,
\begin{equation}
 S\sim S_H
\end{equation}

We shall now move to the calculation of the horizon entropy.
Motivated by the Bekenstein-Hawking \cite{Bekenstein1,Hawking1} formula for black holes, Gibbons and Hawking proposed that
cosmological horizon also posses an entropy
\begin{equation}
 S_H=\frac{A_H k_B}{4L^2_p}
\end{equation}
where $L_p={\sqrt{\frac{\hbar G}{c^3}}}$ is the Planck length and $A_H=4\pi r^2_H$ is the
area of sphere having horizon radius $r_H$\cite{Gibbons1}.
Using the above relation we have calculated the entropy of the Hubble horizon having a radius $r_H=\frac{c}{H}$, where H is
the Hubble parameter as
\begin{equation}
 S_H=\frac{\pi c^2 k_B}{H^2 L^2_p}.
\end{equation}
Substituting H from equation (\ref{eqn:H}), we obtain
\begin{equation}\label{eqn:Hentro}
 S_H={\frac{\pi c^2 k_B}{L^2_p H_0^2 \Omega_\Lambda}}{\tanh^2\left(\frac{3}{2} \sqrt{\Omega_{\Lambda}} H_0 t \right)}
\end{equation}
Assuming the present day Hubble constant $H_0= 70 km/s/Mpc$ and $\Omega_\Lambda=0.7$, the horizon entropy can be estimated as,
$S_H=2.2\times 10^{122} k_B,$ which is close to the value estimated in reference\cite{Egan1}. The horizon entropy, $S_H \to 0$
as $t\to 0$ and $S_H \to 3\times 10^{122} k_B$ asymptotically as $t\to \infty.$

\begin{figure}
\centering
 \includegraphics[width=0.45\textwidth]{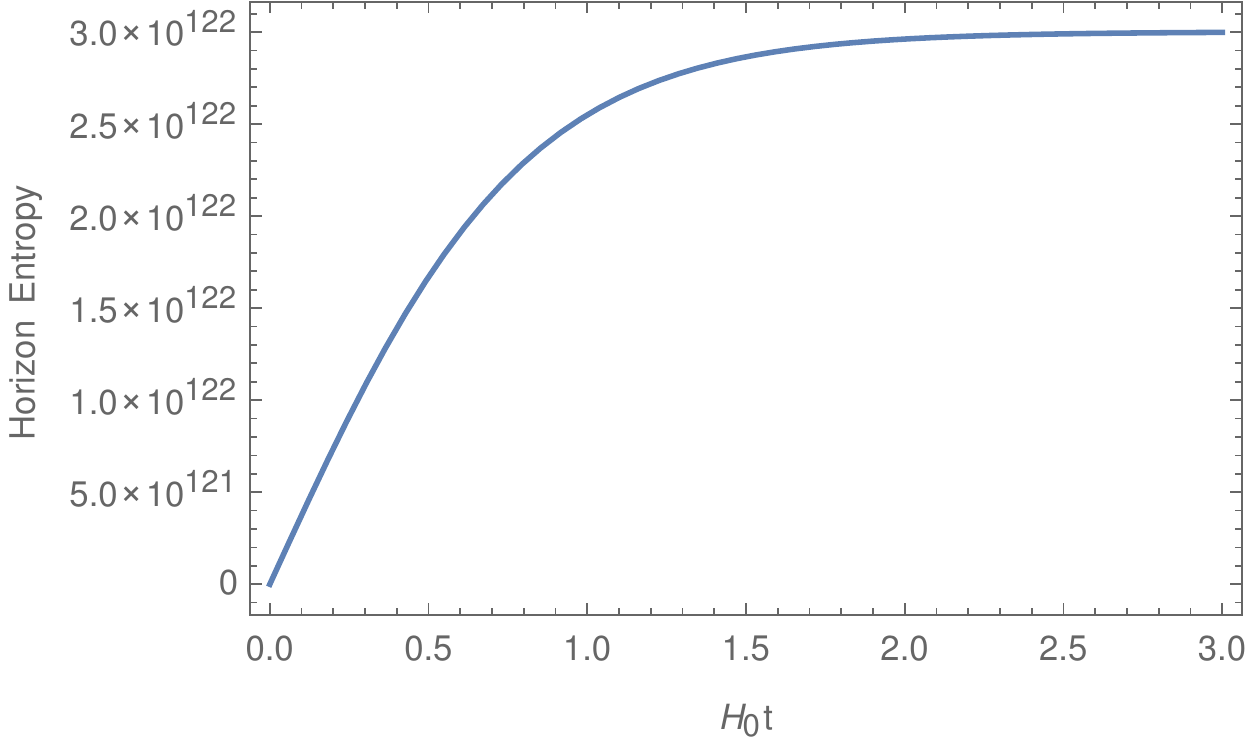}
 \caption{Evolution of the entropy of horizon. The vertical axis shows the horizon entropy in units of $k_B.$}
 \label{fig:entropy1}
\end{figure}

We have plotted the evolution of horizon entropy as the universe expands using equation (\ref{eqn:Hentro})
in figure \ref{fig:entropy1}. From the figure it is seen that the horizon entropy increases and attains a constant maximum value
in the asymptotic limit. That is the total entropy $S \sim S_H,$ never decrease.
Ordinary macroscopic systems naturally evolve to an equilibrium state of maximum entropy. Such systems should be compatible
with the constraints imposed by the generalized second law(GSL). This implies that the total entropy of the system,
$S$ should never decrease, i.e, $S'\geq 0$, where $S'$ is the first derivative of entropy with respect to some relevant
appropriate variable. Further it must be a convex function of the said variable, $S''< 0$, at least at the last stage of evolution.
 If $S''> 0,$ when the variable approaches its final value, an equilibrium state is not achievable and the entropy
will grow unbounded\cite{Diego1}.

As mentioned earlier, since the total entropy, $S$ can be approximated as the horizon entropy $S_H$, one can write the total entropy
similar to equation (\ref{eqn:Hentro}), i.e.,
\begin{equation}
 S \sim {\frac{\pi c^2 k_B}{l^2_p H_0^2 \Omega_\Lambda}}{\tanh^2\left(\frac{3}{2} \sqrt{\Omega_{\Lambda}} H_0 t \right)}
\end{equation}
Now, we have to analyze the behavior of entropy, $S$ and to check whether this universe behave as an ordinary macroscopic system.
Taking the derivative of entropy with respect to the cosmic time $'H_0 t'$, we get
\begin{equation}\label{eqn:s'}
 S'= {\frac{\pi c^2 k_B}{l^2_p H_0^2 \sqrt\Omega_\Lambda}}{3\tanh\left(\frac{3}{2} \sqrt{\Omega_{\Lambda}} H_0 t \right)}
 {sech^2\left(\frac{3}{2} \sqrt{\Omega_{\Lambda}} H_0 t \right)}
\end{equation}
\begin{figure}
\centering
 \includegraphics[width=0.45\textwidth]{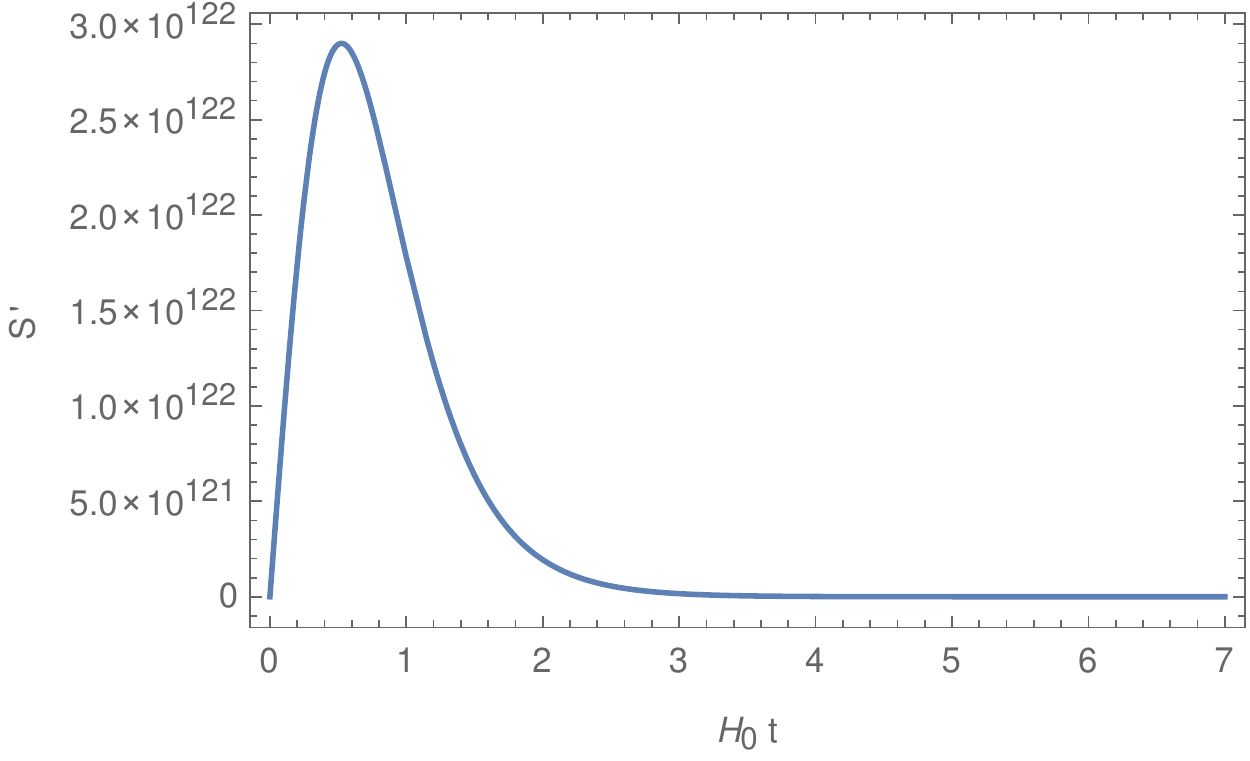}
 \caption{Variation of $S'$ against $H_0 t$. The vertical axis shows the rate of change of entropy with progress of cosmic
 time in units of $k_B.$ }
 \label{fig:S'}
\end{figure}
Figure \ref{fig:S'}  shows the variation of $S'$ with $H_0 t$. The initial increase in $S'$ is
\begin{figure}
\centering
 \includegraphics[width=0.45\textwidth]{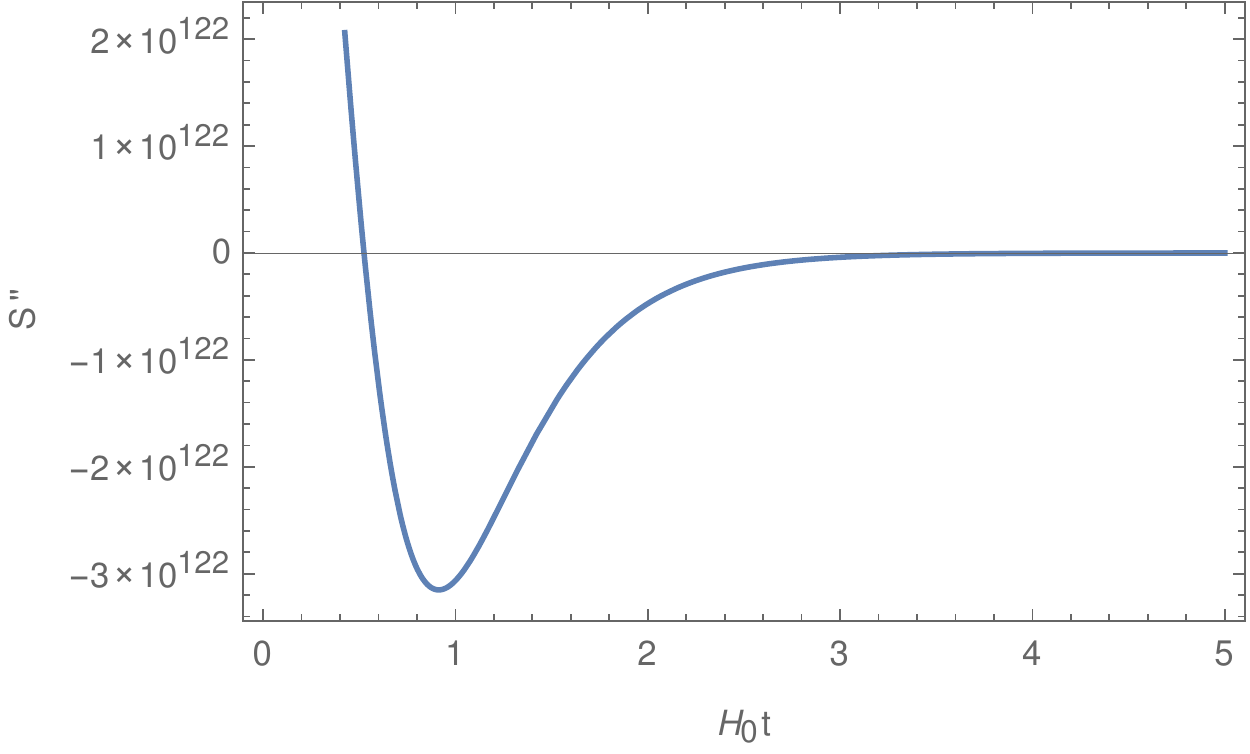}
 \caption{Variation of $S''$ against $H_0 t$. The vertical axis shows the variation of the second derivative of entropy with
 the progress of cosmic time in units of $k_B.$ }
 \label{fig:S''}
\end{figure}
due to monotonic increase in the horizon entropy in the early decelerated stage.
But later $S'$ shows a decrease during the accelerated phase and approaches zero in the final stage when the universe attains
a state of equilibrium, in the asymptotic limit, $t\to\infty.$ This figure guarantees the non-negativity of $S'$ and
thus ensures the consistency with the GSL.

In order to check the maximization of entropy in the final stage, we have to determine the sign of the second derivative of entropy,
$S''$.
Differentiating equation (\ref{eqn:s'}) once more with respect to $'H_0 t'$, one readily get,
\begin{equation}
\begin{aligned}
\begin{split}
 S^{''} = {\frac{9\pi c^2 k_B}{2l^2_p H_0^2}} \left[sech^4\left(\frac{3}{2} \sqrt{\Omega_{\Lambda}} H_0 t \right) \right.- \\
 \left. 2 \tanh^2\left( \frac{3}{2} \sqrt{\Omega_{\Lambda}} H_0 t \right) sech^2\left(\frac{3}{2} \sqrt{\Omega_{\Lambda}} H_0 t\right) \right]
\end{split}
\end{aligned}
\end{equation}
We have plotted the variation of $S''$ against $H_0 t$ in figure \ref{fig:S''}. From the figure it is clear that, $S''< 0$ in
the long run ensuring the convexity of the function. $S''$ approaches 0 from below in the asymptotic limit, when $t\to\infty$.
This figure guarantees that $S''$ will never be positive and hence the entropy will never grow unbounded. Hence this universe
with matter and cosmological constant as the cosmic components behaves as an ordinary
macroscopic system which evolves to an equilibrium state of maximum entropy.

The above results shows the consistency of the standard $\Lambda$CDM model with the equivalency of the holographic equipartition
and the maximum entropy principle. The $\Lambda$CDM model which evolves to a state that satisfies holographic equipartition
also proceeds to a state of maximum entropy and behaves as an ordinary macroscopic system.

\section{Conclusion}

In the present work, we have analyzed the equivalency of the holographic equipartition law and the entropy maximization principle
for an expanding universe. The evolution of the universe can be viewed  as a quest for satisfying the holographic
equipartition. This means that the the rate of change of the Hubble volume is proportional to the discrepancy between the
degrees of freedom on the horizon and in the bulk of the universe. This, in turn, lead us to define the deceleration parameter
as the ratio of the degrees of freedom in the bulk and that on the horizon (equation \ref{eqn:q3}). For a radiation dominated
universe, this implies $N_{surf}=-\epsilon N_{bulk}$. Since $\epsilon=-1$, for radiation, this implies the equality of the
degrees of freedom, $N_{surf}=N_{bulk}.$ However, in this case, the rate of change of Hubble volume
is not zero. In the matter dominated phase the degrees of freedom will never be equal to each other. During late stage of
the evolution, in which dark energy is dominating, the degrees of freedom satisfies the condition, $N_{surf}=\epsilon N_{bulk},$
where $\epsilon=+1$ for dark energy. This again leads to the equality of the degrees of freedom, $N_{surf}=N_{bulk}.$ But
unlike in the radiation dominated phase, here the rate of change of the Hubble volume will become zero.

It has been proved that the universe will behave as an ordinary macroscopic system, which evolves to state of thermodynamic
equilibrium. Thus the entropy $S$ will satisfy the conditions, $\dot S\geq 0$, always and $\ddot S < 0$, in the long run.
If $\ddot S>0$, the entropy may grow unbounded with out attaining an equilibrium state. From the holographic equipartition
we have shown that  the condition $\dot S \geq 0$ leads to $N_{surf}-\epsilon N_{bulk} \geq 0.$ Even though the inequality
(i.e. $\dot S > 0$) is satisfied through out  the evolution of the universe, the limiting condition, the equality, i.e.
$N_{surf}=\epsilon N_{bulk}$ is satisfied only in the dark energy  dominated phase. The condition $\ddot S <0$ is shown to be
equivalent to $\left|\frac{1}{H} \frac{d}{dt}\left(1-\epsilon\frac{N_{bulk}}{N_{surf}}
 \right) \right|> \left(1-\epsilon\frac{N_{bulk}}{N_{surf}} \right)^2.$ This condition is satisfied only in the dark energy
 dominated  phase, where $N_{surf}=\epsilon N_{bulk}$.  It is also shown that the above condition is the same as the
 condition obtained by Pavon and Radicela\cite{Diego1} for entropy maximization in an  expanding universe. This proves the
 equivalency of holographic equipartition law and the maximum entropy principle. In other words, the quest for
 satisfying holographic equipartition can also be interpreted as a tendency to maximize entropy. Thus a system that obeys
 the holographic equipartition  law behaves as an ordinary macroscopic system.

 We have considered the standard $\Lambda$CDM model of the universe and showed that this model is consistent with
 the holographic equipartition. Analyzing the entropy evolution we found that it also proceeds to a maximum entropy state
 and thus behaves as an ordinary macroscopic system. Thus the standard $\Lambda$CDM model supports the equivalency of the holographic equipartition law
 and the maximum entropy principle.
 Both these principles are in general  satisfied by
 quintessence-like dark energy models which in the long run tends towards a de Sitter phase. However, in phantom dark energy
 models  $\ddot S >0$ and cannot satisfy either holographic equipartition or the entropy maximization principle.
 But such models suffer from quantum instabilities\cite{Carrol,Cline} and hence it is difficult to consider
 them as useful dark energy models.

\begin{center}
 {\bf Acknowledgement}
 \end{center}
  We are thankful to the referee for useful suggestions which helped us to improve our manuscript. We are grateful to Prof. M Sabir for the 
  careful reading of the 
  manuscript. We also thank T Padmanabhan for some clarifications. Thanks are also due to IUCAA, Pune for the hospitality during the 
  visit of one of the authors. Krishna P B Acknowledge KSCSTE, Govt. of Kerala for the financial support.



\end{document}